\documentclass[%
 reprint,
nofootinbib,
 amsmath,amssymb,
 aps,
]{revtex4-1}

\usepackage{graphicx,color}
\usepackage{dcolumn}
\usepackage{bm}

\newcommand{\cb}{{\bar{c}}}

\begin{document}

\mbox{}\hfill{LTH 1202}
\vspace{0.7cm}

\title{Two loop calculation of Yang-Mills propagators in the Curci-Ferrari model}

\author{John A. Gracey}
\affiliation{Theoretical Physics Division,
Department of Mathematical Sciences,
University of Liverpool,
P.O. Box 147,
Liverpool,
L69 3BX,
United Kingdom}%

\author{Marcela Pel\'aez}
\affiliation{Instituto de F\'isica, Facultad de Ingenier\'ia, Universidad de la Rep\'ublica, Montevideo, Uruguay}%

\author{Urko Reinosa}
\affiliation{Centre de Physique Th\'eorique (CPHT), CNRS, Ecole polytechnique, Institut Polytechnique de Paris, Route de Saclay, F-91128 Palaiseau, France}%

\author{Matthieu Tissier}%
\affiliation{Sorbonne Universit\'e, CNRS, Laboratoire de Physique Th\'eorique de
la Mati\`ere Condens\'ee, LPTMC, F-75005 Paris, France}%

\date{\today}

\begin{abstract}
  The Landau-gauge gluon and ghost correlation functions obtained in lattice
  simulations can be reproduced
  qualitatively and, to a certain extent, quantitatively if a gluon
  mass is added to the standard Faddeev-Popov action. This has been
  tested extensively at one loop, for the two and three point correlation
  functions of the gluons, ghosts and quarks. In this article, we push the comparison to two loops for
  the gluon and ghost propagators. The agreement between lattice
  results and the perturbative calculation considerably improves. This
  validates the Curci-Ferrari action as a good phenomenological model
  for describing the correlation functions of Yang-Mills theory in the
  Landau gauge. It also indicates that the perturbation theory converges fairly well, in the infrared regime.
\end{abstract}

\maketitle


\section{\label{sec_intro}Introduction}

During the last two decades, there has been an intense activity aimed at 
studying the long-distance properties of the  correlation functions of Quantum 
Chromodynamics (QCD) in the Landau gauge. Nowadays, a consensus has been 
reached in the community. A wide range of approaches (both analytic and 
numerical), concluded that the gluon propagator saturates in the infrared, 
while the ghost propagator diverges. This behavior is consistent with the 
presence of a gluon ``mass", which is found to be of the order of 500 MeV. The 
origin of this mass is, however, still strongly debated. It could be generated 
through nonperturbative effects (captured by truncations of Schwinger-Dyson 
equations \cite{Bloch03,Aguilar04,Boucaud06,Aguilar07,Aguilar08,Boucaud08,RodriguezQuintero10}
or by integrating nonperturbative renormalization-group equations 
\cite{Pawlowski03,Fischer04,Fischer08}), it could result from the generation of
condensates (such as $\langle A^2\rangle$ for instance 
\cite{jag14,jag15,Dudal08}) or could be a consequence of the Gribov ambiguity, 
which invalidates the standard Faddeev-Popov gauge-fixing procedure 
\cite{Serreau:2012cg,Tissier:2017fqf}. From the numerical side, the saturation 
of the gluon propagator is unambiguously seen in lattice simulations 
\cite{Cucchieri_08b,Cucchieri_08,Cucchieri09,Bogolubsky09,Dudal10}.

Understanding the origin of this mass is of great relevance to the field, but
remains a difficult task. A more humble program consists in studying to what 
extent the long-distance behavior of QCD is related to the presence of this 
mass. One way of addressing this question is to minimally extend the Landau 
gauge-fixed QCD Lagrangian by means of a mass term for the gluons, added on 
phenomenological grounds. This starting point corresponds to the Curci-Ferrari 
model, in the limit of vanishing gauge parameter. In a series of articles 
\cite{Tissier:2010ts,Tissier:2011ey,Pelaez:2013cpa,Pelaez:2014mxa,Pelaez:2015tba,Pelaez:2017bhh, Reinosa:2017qtf}, 
the $2$- and $3$-point correlation functions of gluons, quarks and ghosts were 
computed at leading (one loop) order in perturbation theory and the results 
were compared to available lattice simulations.\footnote{A gluonic mass operator has also been considered by F.~Siringo and collaborators, within the general setting of $R_\xi$-gauges \cite{Siringo:2015wtx,Siringo:2018uho,Siringo:2019qwx}. However, contrary to the present approach, the working hypothesis in this case is that the Faddeev-Popov action is not modified by the presence of Gribov copies. The gluonic operator is formally added and subtracted to the Faddeev-Popov action as a way to reorganize the standard perturbative expansion of the model, while curing its bad features in the infrared. The value of the gluonic mass is typically fixed by some optimization criterion.} The overall picture which 
emerges is the following. In the quenched approximation (or Yang-Mills theory),
where the fluctuations of the quarks are neglected, the lattice results can be 
estimated with a maximal error of 10-20$\%$ on the whole range of available 
momenta. The model is therefore very predictive, since many features can be 
reproduced with only {\em one} phenomenological parameter: the gluon mass. 
These results are surprising at first sight because the infrared regime of QCD 
is reputed to be nonperturbative. The apparent paradox can be solved by 
observing that the coupling constant deduced from lattice simulations (see eg 
Ref.~\cite{Cucchieri:2008qm}) and derived from analytic calculations 
\cite{Dudal:2012zx} remains finite and quite mild in the whole range of 
momenta. This is at odds with the result obtained in standard perturbation 
theory, where the coupling constant is found to diverge at an energy scale of 
the order of few hundreds of MeV. Within the Curci-Ferrari model, it is 
explicitly seen \cite{Tissier:2011ey} that the mass term regularizes the 
infrared properties of the theory, which does not experience a Landau pole. 
When the quark dynamics is taken into account, the agreement between lattice 
simulations and the one-loop results is worse. This is related to the fact that
the quark-gluon interaction is up to three times bigger than the $3$-gluon 
interaction (the ghost-gluon interaction being of the order of the 
latter).\footnote{We stress that the behaviour of the running coupling constant is universal in the deep
ultraviolet but the interaction in different channels can (and does) differ in 
the infrared limit.} In this situation, it proved useful to treat the 
ghost-gluon sector perturbatively, while treating the matter sector using an 
expansion in the inverse of the number of colors. This strategy enables the 
description of the chiral transition within a systematic expansion controlled 
by two small parameters \cite{Pelaez:2017bhh}.

The aim of this article is to further test the convergence of the theory. It is
particularly important, for our whole project to understand:
\begin{itemize}
\item to which {extent} perturbation theory converges in the quenched limit. 
More precisely, it is important to evaluate the contribution of higher loops to
some observables
\item if the theory converges, does it converge to results close to those of 
the true Yang-Mills theory. The issue here is to test the validity of the 
phenomenological model.
\end{itemize}
We concentrate here on the gluon and ghost propagators in the quenched 
approximation which are the simplest correlation functions to compute and for 
which we have the cleanest lattice data. 

The rest of the article is organized as follows. In Section \ref{sec_CF}, we 
recall the model and describe the renormalization scheme that we use. We give 
some details of the two-loop calculation in Section \ref{sec_Comp}. In Section 
\ref{sec_analytic}, we discuss how analytic results can be obtained in some 
momentum configurations. We finally compare the perturbative results to lattice
data in Section \ref{sec_res}.

\section{Curci-Ferrari model}
\label{sec_CF}
Based on the phenomenological considerations given above, we use as a starting 
point the following Lagrangian density
\begin{equation}
  \label{eq_lagrang}
  \mathcal L=\frac 14 (F_{\mu\nu}^a)^2+\partial _\mu\overline c^a(D_\mu
  c)^a+ih^a\partial_\mu A_\mu^a+\frac {m^2}2 (A_\mu^a)^2\,.
\end{equation}
{The covariant derivative} $(D_\mu c)^a=\partial_\mu c^a+g f^{abc}A_\mu^b c^c$ 
{ and the field strength} 
$F_{\mu\nu}^a=\partial_\mu A_\nu^a-\partial_\nu A_\mu^a+gf^{abc}A_\mu^bA_\nu^c$
are expressed in terms of the coupling constant $g$ and the Latin indices 
correspond to the SU$(N_c)$ gauge group. The Lagrangian~(\ref{eq_lagrang}) 
corresponds to a particular case of the Curci-Ferrari model~\cite{Curci76}, 
obtained in the limit of vanishing gauge parameter. At tree level, the gluon 
propagator is massive and transverse in momentum space, which ensures that the 
model is renormalizable. Note that the mass term is introduced at the level of 
the gauge-fixed theory. If instead we modifed the unfixed theory, we would 
obtain a longitudinal propagator which does not decrease in the ultraviolet and
the theory would not be renormalizable. We refer the reader to Ref. 
\cite{Tissier:2011ey} for a more detailed account of this model, including its 
symmetries.

The theory is regularized in $d=4-2\epsilon$ dimensions. It is renormalized by introducing renormalized coupling constant, mass 
and fields, which are related to the bare ones (that we denote now with the 
subscript ``$B$'') by including multiplicative renormalization factors: 
\begin{align}
A_B^{a\,\mu}= \sqrt{Z_A} A^{a\,\mu},&\hspace{.5cm} 
c_B^{a}= \sqrt{Z_c} c^{a},\hspace{.5cm}
\bar c_B^{a}= \sqrt{Z_c} \bar c^{a},\hspace{.5cm} \nonumber\\
\lambda_B&= Z_g^2 \lambda, \hspace{.5cm} m_B^2= Z_{m^2} m^2,\label{eq:rescaling}
\end{align}
with $\lambda=\frac{g^2 N_c}{16\pi^2}$.\footnote{Note that, at order $\lambda$, 
the renormalization factors need to be expanded up to order $\epsilon^1$. This 
is because, they appear in one-loop diagrams which diverge as $1/\epsilon$. The
combination of these divergences together with the $\epsilon^1$ of the 
renormalization factors produce finite contributions (of order $\lambda^2$) 
that should not be forgotten in the two-loop calculation considered in this 
work.} The renormalization factors are defined by choosing the value of propagators 
and vertices at a given scale $\mu$.
For the gluon propagator and the ghost dressing functions, we choose
\begin{align}
& G^{-1}(p=\mu)=m^2+\mu^2, \hspace{.4cm}F(p=\mu)=1\,.
\end{align}
We use the Taylor scheme to fix the renormalization of the coupling constant. 
In this scheme, the coupling constant is defined as the ghost-gluon vertex with
a vanishing antighost momentum. This leads to the following relation between 
renormalization factors
\begin{equation}
 \label{eq_taylor}
 Z_g \sqrt{Z_A} Z_c=1\,.
\end{equation}
Finally, we use the non-renormalization theorem for the divergent part of the 
gluon mass \cite{jag12,jag13,Dudal02,Wschebor07,Tissier08}:
\begin{align}
\label{IScondition}
&Z_{m^2} Z_A Z_c=1\,,
\end{align}
and extend this relation to the finite parts. The four previous constraints 
define the Infrared Safe (IS) scheme \cite{Tissier:2011ey}.\footnote{A similar nonrenormalization theorem for the Gribov mass parameter was derived in \cite{maggiore94,dudal05}}

We obtain the flow of the coupling constant, the gluon mass and the anomalous 
dimensions by computing:
\begin{align}
\label{Eq:RGflow}
\beta_\lambda(\lambda,m^2)&=\mu\frac{d\lambda}{d\mu}\Big|_{\lambda_B, m^2_B},\nonumber\\
\beta_{m^2}(\lambda,m^2)&=\mu\frac{dm^2}{d\mu}\Big|_{\lambda_B, m^2_B},\nonumber\\
\gamma_A(\lambda,m^2)&=\mu\frac{d\log Z_A}{d\mu}\Big|_{\lambda_B, m^2_B}\,,\nonumber\\
\gamma_c(\lambda,m^2)&=\mu\frac{d\log Z_c}{d\mu}\Big|_{\lambda_B, m^2_B}\,.
\end{align}
Thanks to the non-renormalization theorems, see Eqs. (\ref{eq_taylor}) and 
(\ref{IScondition}), the anomalous dimensions are easily related to the beta 
functions as:
\begin{align}
\label{eqgama}\gamma_A(\lambda,m^2)&= 2\frac{\beta_{m^2}}{m^2}-\frac{\beta_\lambda}{\lambda},\\
\label{eqgamc}\gamma_c(\lambda,m^2)&=\frac{\beta_\lambda}{\lambda}-\frac{\beta_{m^2}}{m^2}.
\end{align}

We can then use the renormalization-group (RG) equation for the vertex function with $n_A$ gluon legs 
and $n_c$ ghost legs:
\begin{equation}
\begin{split}
\Big( \mu \partial_\mu -\frac 1 2 &(n_A \gamma_A+n_c \gamma_c)\\&+\beta_\lambda 
\partial_{\lambda}+
\beta_{m^2}\partial_{m^2}\Big)\Gamma^{(n_A,n_c)}=0\,,
\end{split}
\end{equation}
to relate these functions at different scales:
\begin{equation}
\label{eq_int_RG}
\begin{split}
&\Gamma^{(n_A,n_c)}(\{p_i\},\mu,\lambda(\mu),m^2(\mu))=z_A(\mu;\mu_0)^{
n_A/2}\\
&\times z_c(\mu;\mu_0)^{
n_c/2}\Gamma^{(n_A,n_c)}(\{p_i\},\mu_0,\lambda(\mu_0),m^2(\mu_0))\,,
\end{split} 
\end{equation} 
where $\lambda(\mu)$ and $m^2(\mu)$ are obtained by integration of the
beta functions with initial conditions given at some scale $\mu_0$ and where:  
\begin{equation}
\label{eq_def_z_phi}
\begin{split}
\log z_A(\mu;\mu_0)&=\int_{\mu_0}^\mu\frac
    {d\mu'}{\mu'}\gamma_A\left(\lambda(\mu'),m^2(\mu')\right),\\ \log
    z_c(\mu;\mu_0)&=\int_{\mu_0}^\mu\frac
    {d\mu'}{\mu'}\gamma_c\left(\lambda(\mu'),m^2(\mu')\right).
\end{split}
\end{equation}

In order to avoid large logarithms we choose the renormalization group scale 
$\mu=p$ in Eq.~(\ref{eq_int_RG}). We thus obtain
\begin{align}
G(p;\mu_0)&=\frac{z_A(p;\mu_0)}{p^2+m^2(p)}\,,\nonumber\\
F(p;\mu_0)&=z_c(p;\mu_0)\,.
\end{align}
By using the non-renormalization theorems (\ref{eqgama})-(\ref{eqgamc}), the 
gluon propagator and the ghost dressing function are readily deduced from the 
running parameters:
\begin{align}
\label{propsInIS}
{ G(p;\mu_0)} &=\frac{\lambda(\mu_0)}{m^4(\mu_0)}\frac{m^4(p)}{\lambda(p)}\frac{1}{p^2+m^2(p)}\,,\nonumber\\
{ F(p;\mu_0)} &=\frac{m^2(\mu_0)}{\lambda(\mu_0)}\frac{\lambda(p)}{m^2(p)}\,.
\end{align}
One advantage of the IR scheme is that the propagators at some momentum scale 
are algebraically related to the running mass and coupling constant, evaluated 
at the same momentum scale.\\

\vglue10mm

\section{Computational details}
\label{sec_Comp}

We devote this section to detailing the evaluation of the underlying Feynman 
graphs contributing to the gluon and ghost propagators in the Landau gauge of 
Yang-Mills theory when there is a non-zero gluon mass. To achieve this, we have 
constructed an automatic routine which computes the $2$-point functions using 
state of the art Feynman diagram evaluation procedures. The starting point is 
the construction of the Feynman graphs for each Green's function and for this 
we used the graph generator {\sc Qgraf}, \cite{jag1}. Since there is a non-zero
gluon mass, we have been careful to include graphs involving gluon snails in the
language of \cite{jag1}. Ordinarily, the gluon is regarded as massless. So 
graphs where the quartic gluon vertex is included on a gluon propagator with 
two legs contracted are omitted as these would be zero in dimensional 
regularization. With a non-zero $m$, such graphs will give contributions and are
included at one and two loops. Also, omitting them would lead to inconsistencies
with the gluon mass renormalization. In total, there are $16$ two loop graphs 
for the gluon $2$-point function and $6$ for the ghost case. At one loop, the 
respective numbers are $3$ and $1$. Once the graphs are generated, the 
electronic representation is converted into the notation of the symbolic 
manipulation language {\sc Form}, \cite{jag2,jag3}. It is the most suitable 
tool to handle the large tedious amounts of internal algebra. With a non-zero
mass, it is not possible to use established diagram evaluation packages and, 
therefore, we have resorted to implementing the Laporta algorithm, \cite{jag4}, 
for the computation. To apply it, each Green's function needs to be written as
a sum of scalar integrals. This is achieved by writing all the scalar products 
in the form of the propagators. For the
gluon propagator, one has to first project out the transverse and longitudinal 
components. To two loop order, this procedure is straightforward as the number 
of independent scalar products of internal and external momenta equals the 
total number of propagators in the one and two loop integral families in the 
syntax of the Laporta technique, \cite{jag4}. These are illustrated in 
Fig.~\ref{fig1cr} where the one loop and two loop $d$-dimensional integrals are 
defined to be 

\begin{widetext}
\begin{eqnarray}
I_{1 m_1 m_2}(n_1,n_2) &=& \int_{k} \frac{1}{[k^2+m_1^2]^{n_1} 
[(k-p)^2+m_2^2]^{n_2}}\,,\nonumber \\
I_{m_1 m_2 m_3 m_4 m_5}(n_1,n_2,n_3,n_4,n_5) &=&
\int_{kl} \frac{1}{[k^2+m_1^2]^{n_1} [l^2+m_2^2]^{n_2} [(k-p)^2+m_3^2]^{n_3}
[(l-p)^2+m_4^2]^{n_4} [(k-l)^2+m_5^2]^{n_5}}\,,\nonumber \\
\label{intfam}
\end{eqnarray}
with $\int_k$~$=$~$\int d^dk/(2\pi)^d$ and the variables $n_i$ are integers.

\begin{center}
{\begin{figure}[ht]
\begin{center}
\includegraphics[width=10.0cm,height=2.6cm]{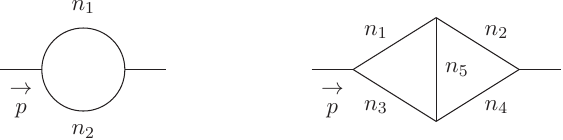}
\end{center}
\caption{One and two loop integral families.}
\label{fig1cr}
\end{figure}}
\end{center}
\end{widetext}

In the decomposition of each of the Green's functions into scalar integrals,
the propagator powers can be larger than unity. Equally, one can have integrals
where the scalar products of momenta, when rearranged, exceed the number of
corresponding denominator factors. So, in Eq.~(\ref{intfam}), the propagator 
powers can be negative. Such integrals are readily accounted for in the Laporta
construction. The notation is that the powers of the propagtors, which can 
therefore be negative or zero as well as positive, appear in the arguments of
the respective functions representing the two integral families. With each, we
have to allow for all possible distributions of a non-zero mass of each
propagator. Therefore, the masses $m_i$ take values in the set $\{0,m\}$ since
the ghosts are massless and there is a transverse tensor in the 
gluon propagator. To reflect this within the notation, we append subscripts to 
$I$ in the definitions in (\ref{intfam}) which take values in $\{0,1\}$, where 
$1$ corresponds to the mass being non-zero on the respective propagator. For 
orientation, we provide several examples which are
\begin{widetext}
\begin{eqnarray}
I_{110}(n_1,n_2) &=& \int_{k} \frac{1}{[k^2+m^2]^{n_1} [(k-p)^2]^{n_2}}\,, 
\nonumber \\
I_{00000}(n_1,n_2,n_3,n_4,n_5) &=& \int_{kl} 
\frac{1}{[k^2]^{n_1} [l^2]^{n_2} [(k-p)^2]^{n_3}
[(l-p)^2]^{n_4} [(k-l)^2]^{n_5}}\,,\nonumber \\
I_{11101}(n_1,n_2,n_3,n_4,n_5) &=& 
\int_{kl} \frac{1}{[k^2+m^2]^{n_1} [l^2+m^2]^{n_2} [(k-p)^2+m^2]^{n_3}
[(l-p)^2]^{n_4} [(k-l)^2+m^2]^{n_5}}\,.
\end{eqnarray}
\end{widetext}
Once each of the Green's function is written in terms of these two core
integral structures, the Laporta algorithm is applied. We have chosen to use
the {\sc Reduze} version, \cite{jag5,jag6}, and each of the integrals is
converted to the unique Laporta labelling, \cite{jag4}. Using {\sc Reduze}, we
have created a database of the relations for the required scalar integrals, for
all possible mass configurations. These are then solved algebraically within
{\sc Reduze} in order to relate all integrals to a basic set of what are known 
as master integrals. In the case of the present problem, there are $2$ one loop 
masters and $31$ two loop ones. The latter total includes those cases which are
the disjoint product of one loop masters. In addition to the integral family 
topologies of Fig.~\ref{fig2cr}, there are several additional master 
topologies which are illustrated in Fig.~\ref{fig2cr}. The large number of 
masters is due to the different ways the topologies can be decorated with 
non-zero mass. In that file, we use the Laporta labelling, 
\cite{jag4}, and to assist with this we note  
\begin{widetext}
\begin{eqnarray}
I_{1 a b}(n_1,n_2) &=& {\tt int1ab ( t, id, r, s, n_1, n_2 )}\,, \nonumber \\ 
I_{a b c d e}(n_1,n_2,n_3,n_4,n_5) &=&
{\tt intabcde ( t, id, r, s, n_1, n_2, n_3, n_4, n_5 )}\,.
\label{intdef}
\end{eqnarray}
The first four entries of each integral in the {\sc Form} output are the unique
internal labels required by the {\sc Reduze} version of the Laporta algorithm. 
In particular, {\tt id} labels the sector the integral belongs to uniquely. It 
is defined from the different ways the various lines can appear in each of the 
Fig.~\ref{fig1cr} integral families. This includes the case where there are 
no lines which is known as a zero sector. At one loop, there are four sectors 
but $32$ at two loops for each possible mass configuration. The total number of
independent lines in a sector is {\tt t} irrespective of what their powers are.
The sum of the propagator powers is {\tt r} while the sum of numerator 
propagators is {\tt s}. Several of the masters have non-unit powers which is an
established feature of master bases. An example where one can be related to 
other integrals is
\begin{equation}
I_{10011}(1,-1,0,1,1) ~=~ 
\frac{1}{3} [ p^2 - 3 m^2 ] I_{10011}(1,0,0,1,1) ~+~ I_{11000}(1,1,0,0,0)\,,
\end{equation}
where $p$ is the external momentum. In the Supplemental material \cite{jag7}, we give the 2-loop expressions for the gluon and ghost 2-point vertices expressed in terms of the integrals {\tt int1ab} and {\tt intabcde} defined above. We have used the more general 
gluon-ghost vertex of the gauge fixing of Curci-Ferrari, \cite{Curci76} which involves an extra 
parameter $\beta$ for 
any future investigations into other gauges but our results focus exclusively 
throughout on the standard Faddeev-Popov case which is $\beta$~$=$~$1$. Finally, the transverse and longitudinal parts of the 2-point gluon vertex are encoded in the parts proportional to {\tt long} and {\tt trans} resppectively. For 
completeness, we note that the Feynman rule for the gluon-ghost vertex used is 
\begin{equation}
\langle A^a_\mu(p_1) \bar{c}^b(p_2) c^c(p_3) \rangle ~=~
-~ \frac{i}{2} g f^{bac} \left[ [1+\beta] p_{2\,\mu} - [1-\beta] p_{3\,\mu}
\right]\,.
\end{equation}
where $g$ is the usual gauge coupling constant.

\begin{center}
{\begin{figure}[ht]
\begin{center}
\includegraphics[width=9.0cm,height=4.8cm]{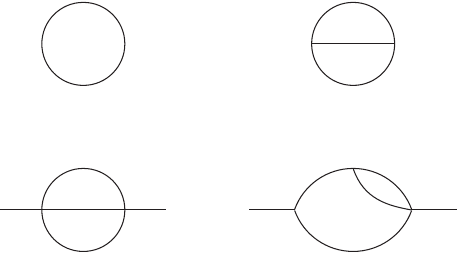}
\end{center}
\caption{Additional one and two loop master topologies.}
\label{fig2cr}
\end{figure}}
\end{center}
\end{widetext}

The main duty of the Laporta algorithm is to convert the evaluation of a
Green's function into a small set of master integrals. The next stage in the
procedure is the determination of these integrals. How one achieves this is
dependent on the particular problem of interest. Here, we wish to plot the 
propagators over the whole momentum range and compare with the lattice gauge 
theory computation of the same quantities. Therefore, we either need the 
explicit analytic form of all the master integrals as a function of $m$ and the
external momentum or instead a numerical tabulation of each  master integral. 
In the former case, 
while there has been large amount of investment in achieving this for two loop 
self-energy graphs where the distribution of masses corresponds to integrals 
which can appear in the Standard Model at large, such as in \cite{jag9}, 
integrals like $I_{11111}(1,1,1,1,1)$ are not known explicitly analytically. 
Various known cases have been noted in \cite{jag9}. So, instead, we have resorted
to a numerical analysis and made extensive use of the {\sc Tsil} package, 
\cite{jag10}, which is written in C. This has been designed with the Laporta 
algorithm in mind as it provides the necessary tools to numerically evaluate 
two loop self-energy graphs with all possible mass configurations. Moreover, it 
is comprehensive and complete in the sense that we have not taken the route of 
trying to consolidate results for various integrals from different sources. 
This would have the added complication of needing to reconcile different 
conventions. As an aid to converting between the {\sc Reduze} conventions of 
(\ref{intfam}) and (\ref{intdef}) and the integral definitions of {\sc Tsil}, the mapping between the two is 
\begin{equation}
I_{1x0}(1,0) ~=~ {\mathbf A}(x)\,,~~
I_{1xy}(1,1) ~=~ {\mathbf B}(x,y)\,, 
\end{equation}
for the one loop integrals of {\sc Tsil}. As {\sc Tsil} accommodates the most
general mass configuration, we adapt our definitions (\ref{intfam}) and use $x$,
$y$, $z$, $u$ and $v$ as subscript labels corresponding to the different 
masses. These parameters are used in {\sc Tsil} and correspond to the general
masses $m_i^2$ of our notation in (\ref{intfam}) although we have only one
mass in our problem. At two loops, the corresponding relations are 
\begin{eqnarray}
I_{xy00z}(1,1,0,0,1) &=& {\mathbf I}(x,y,z)\,, \nonumber \\
I_{x00yz}(1,0,0,1,1) &=& {\mathbf S}(x,y,z)\,, \nonumber \\
I_{x00yz}(2,0,0,1,1) &=& {\mathbf T}(x,y,z)\,, \nonumber \\
I_{yux0z}(1,1,1,0,1) &=& {\mathbf U}(x,y,z,u)\,, \nonumber \\
I_{yux0z}(2,1,1,0,1) &=& {\mathbf V}(x,y,z,u)\,, \nonumber \\
I_{xyzuv}(1,1,1,1,1) &=& {\mathbf M}(x,y,z,u,v)\,.
\end{eqnarray}

One useful aspect of the {\sc Tsil} package used in this work is that it { correctly} isolates the poles with respect to $\epsilon$ { for each master integral} ahead of the numerical evaluation of the finite part. Moreover, the residues of the $\frac{1}{\epsilon}$ and $\frac{1}{\epsilon^2}$ 
poles are determined analytically rather than numerically. This provides an important check on both the Laporta reduction and the {\sc Tsil} 
implementation. The divergent part for each $2$-point function is already known
analytically in the case of a massless gluon. This, therefore, has to be 
recovered and we note that this is indeed the case when the $m$~$\to$~$0$ limit
is taken. Moreover, we correctly recover the divergent part of the gluon mass 
renormalization to two loops when $m$~$\neq$~$0$. Therefore, the correct 
$\overline{\mbox{MS}}$ renormalization constants emerge from our full 
contruction of the Green's functions. The main benefit of {\sc Tsil} 
\cite{jag10} is that if the finite part of a master is unknown then it is 
evaluated numerically and we note that version $1.41$ was used for our 
computations. 

 \section{Analytic results}
\label{sec_analytic}

After extracting the divergences of the various master integrals according to 
Ref.~\cite{jag10} and implementing the IR safe renormalization conditions, the 
decomposition of the inverse gluon propagator and of the inverse ghost dressing
function into master integrals can be written formally as
\begin{eqnarray}
G^{-1}(p) & = & Z_1\,p^2+Z_2 m^2+\sum_{{\cal I}\in {\cal M}} {\cal R}_G({\cal I})\,{\cal I}\,,\label{eq:decomp_G}\\
F^{-1}(p) & = & Z_3+\sum_{{\cal I}\in {\cal M}} {\cal R}_F({\cal I})\,{\cal I}\,,\label{eq:decomp_F}
\end{eqnarray}
where the $Z_i$ are related to the finite parts of the renormalization factors,
$\sum_{{\cal I}\in{\cal M}}$ represents the sum over the finite master 
integrals and the coefficients ${\cal R}_{G}({\cal I})$ and 
${\cal R}_{F}({\cal I})$ are rational functions of the external momentum $p$ 
which depend on the considered integral. The finite master integrals are 
defined in Ref~\cite{jag10} and are denoted $A_0$, $B_0$, $I_0$, $S_0$, $T_0$, 
$U_0$, $V_0$ and $M_0$. The sum also involves products of bilinears in $A_0$ 
and/or $B_0$ and also linear terms involving the order $\epsilon^1$ 
contributions to the one-loop master integrals, denoted respectively $A_1$ and 
$B_1$.\footnote{The origin of those terms is again that the order $\epsilon^1$ 
contributions multiply factors of order $1/\epsilon$. For completeness, we also
mention that the renormalization procedure generates linear terms involving 
$m^2$-derivatives of one-loop master integrals. However, using dimensional 
analysis, these derivatives can always be expressed in terms of the master 
integrals themselves.} Similar decompositions hold for the $\beta$- and 
$\gamma$-functions. The explicit expressions for the renormalized 2-point vertex in an arbitrary scheme, expressed in terms of these finite master integrals is given in the Supplemental Material \cite{jag7}. We also included in this Supplemental Material the two-loop expressions for the renormalization factors and the $\gamma$ functions.

In principle, the above decompositions can be evaluated directly using the 
{\sc Tsil} library. In practice however, certain ranges of momenta, including 
the ultraviolet regime ($p\gg m$), the infrared regime ($p\ll m$), and also the
vicinity of $p=p_*\equiv\sqrt{2}m$, require a more analytic treatment. The 
reason is that the decompositions (\ref{eq:decomp_G}) and (\ref{eq:decomp_F}) 
are not well conditioned in those ranges of momenta because the expected 
behavior of $G^{-1}(p)$ and $F^{-1}(p)$ as functions of $p$ emerges only as the
result of cancellations among the various terms of the decomposition.

For instance, we find that some of the terms in the decomposition diverge 
artificially as $p$ approaches $p_*$, whereas we expect $G^{-1}(p)$ and 
$F^{-1}(p)$ to be regular for any value of the Euclidean momentum. A more 
detailed analysis reveals that the residue of the potential pole at $p=p_*$ is 
proportional to
\begin{eqnarray}
& & 11-8 \frac{A_0(m^2)}{m^2}+2\frac{A_0(m^2)^2}{m^4}-4\frac{A_0(m^2)}{m^2}B_0(m^2,m^2;p^2_*)\nonumber\\
& & \hspace{0.4cm}-\,2B^2_0(m^2,m^2;p^2_*)+4B_0(m^2,m^2;p^2_*)\nonumber\\
& & \hspace{0.4cm}+\,\frac{4}{m^2}S_0(m^2,m^2,m^2;p^2_*)\,,
\end{eqnarray} 
both for the gluon propagator and for the ghost dressing function. All these 
integrals can be computed analytically \cite{jag10} and we find that the 
residue vanishes, as it should. This is a non-trivial check of our 
decompositions into master integrals.

Similarly, we find that the individual terms in (\ref{eq:decomp_G}) can grow up
to $p^6$ (up to logarithmic corrections) in the UV, and those in 
(\ref{eq:decomp_F}) can grow up to $p^4$, whereas on general grounds, we expect
\begin{equation}
\lim_{p\to\infty}\frac{G^{-1}(p)}{|p|^3}=0 \,\, \mbox{and} \lim_{p\to\infty}\frac{F^{-1}(p)}{|p|}=0\,.
\end{equation}
In the infrared, the various terms can grow up to $1/p^4$ in the IR, whereas we
expect the $G^{-1}(p)$ and $F^{-1}(p)$ to be regular. In order to check that 
the correct behavior emerges after summing over the master integrals we have 
used analytic asymptotic expansions for the master integrals that we derived 
using the algorithms described in \cite{Davydychev:1993pg,Davydychev:1992mt}. 
The algorithm in the infrared does not apply to certain cases but fortunately 
the master integrals in those cases are known exactly \cite{jag10}. Another possible strategy could be to try to use the algorithms developped in \cite{Berends:1994sa,Berends:1996gs}. In the 
Supplemental Material \cite{jag7}, we provide the infrared and ultraviolet expansions of 
all the master integrals needed in our calculation up to order  $p^2$.\footnote{Some of these master integrals are known analytically. We only give the expansions of those for which no analytic form is known.}

Using these expansion, we find for instance that the potentially dangerous 
$1/p^4$ contributions in the infrared regime are proportional to
\begin{equation}
\propto \lambda^2\Big[A_0(m^2)+m^2B_0(m^2,0;p^2=0)\Big]^2\frac{1}{p^4}
\end{equation}
and thus cancel identically since the combination between brackets vanishes. 
Similarly, the dangerous $1/p^2$ contributions cancel owing to an identity 
between the finite two-loop master integrals that involves the following 
relation
\begin{equation}
0=\pi^2-36\,{\rm Li}_2(e^{-i\pi/3})-54i{\rm Cl}_2\left(\frac{2\pi}{3}\right)
\end{equation}
between Clausen function and the dilogarithm. 

We have used the UV/IR asymptotic expansions of the master integrals not only 
to check that the expected leading behaviors of $G^{-1}(p)$ and $F^{-1}(p)$ are
retrieved in the ultraviolet and the infrared, but also to replace when needed 
the numerical evaluation of these functions by a controlled analytic expansion 
to arbitrary order. In particular to leading order, we find that the UV 
behavior are given by
\begin{eqnarray}
\gamma_A & = & -\frac{13}{3}\lambda-\frac{85}{6}\lambda^2+{\cal O}\left(\frac{m^2}{\mu^2}\ln\frac{m^2}{\mu^2}\right),\\
\gamma_c & = & -\frac{3}{2}\lambda-\frac{17}{4}\lambda^2+{\cal O}\left(\frac{m^2}{\mu^2}\ln\frac{m^2}{\mu^2}\right),
\end{eqnarray}
from which one recovers the two-loop universal $\beta_\lambda$ function. The 
expansion in the infrared regime leads, for the IR safe scheme, to
\begin{align}
  \gamma_A  =&  \lambda\left(\frac{1}{3}-\frac{217\bar\mu^2}{180m^2}\right)+\frac{\lambda^2\bar\mu^2}{m^2}\left[\frac{38687}{25920}-\frac{37 }{48} \zeta(2)\right.\\&\left.+\frac{3647 }{288}S_2-\frac{179}{180}\log\left(\frac{\bar\mu}{m}\right)+\frac{13}{36}\log^2\left(\frac{\bar\mu}{m}\right)\right]\nonumber\\&+\mathcal O(\lambda^3,(\bar\mu/m)^4)\,,\nonumber\\
  \gamma_c=&\frac{\lambda\bar \mu^2}{m^2}\left[-\frac 5{12}+\log\left(\frac{\bar\mu}{m}\right)\right]+\frac{\lambda^2\bar \mu^2}{m^2}\left[-\frac{4295}{576}\right.\\&\left.+\frac{5 
   }{12}\zeta(2)+\frac{459 }{16}S_2+\frac{1}{6} \log \left(\frac{\bar \mu}{m}\right)\right]\nonumber\\&+\mathcal O(\lambda^3,(\bar\mu/m)^4)\,,\nonumber
\end{align}
where the constant $S_2=2 \sqrt 3 /9\, \text{Cl}_2(2\pi/3)\simeq 0.2604341$. It
is remarkable that the low-momentum behavior of $\gamma_A$ is completely 
governed by the 1-loop result. 

As a further crosscheck of our calculation, we have used the symmetries of the 
model. In particular, the Curci-Ferrari model possesses a (non-nilpotent) version of Becchi-Rouet-Stora-Tyutin (BRST) symmetry. Together with the equation of motion for the antighost 
field, this symmetry implies a relation between the ghost dressing function and
the longitudinal component of the vertex function $\Gamma^{(2)}_{AA}$, namely 
\begin{equation}
\Gamma^{(2)}_{AA,L}(p) F^{-1}(p)=m^2,
\end{equation}
see \cite{Tissier:2011ey} for more details. We have checked that this identity 
is fulfilled by our expressions up to the relevant order of accuracy.

\section{Results}
\label{sec_res}
We are now ready to compute explicitly the gluon and ghost propagators at two
loop order. We shall do so in four dimensions for the $SU(2)$ and $SU(3)$ gauge
groups and compare our results with lattice data as well as with previously 
obtained one loop results. However, before embarking in this discussion, we 
first describe the general properties of the RG flow.

\subsection{General properties of the renormalization-group flow}

\begin{figure}
\includegraphics[width=0.45\textwidth]{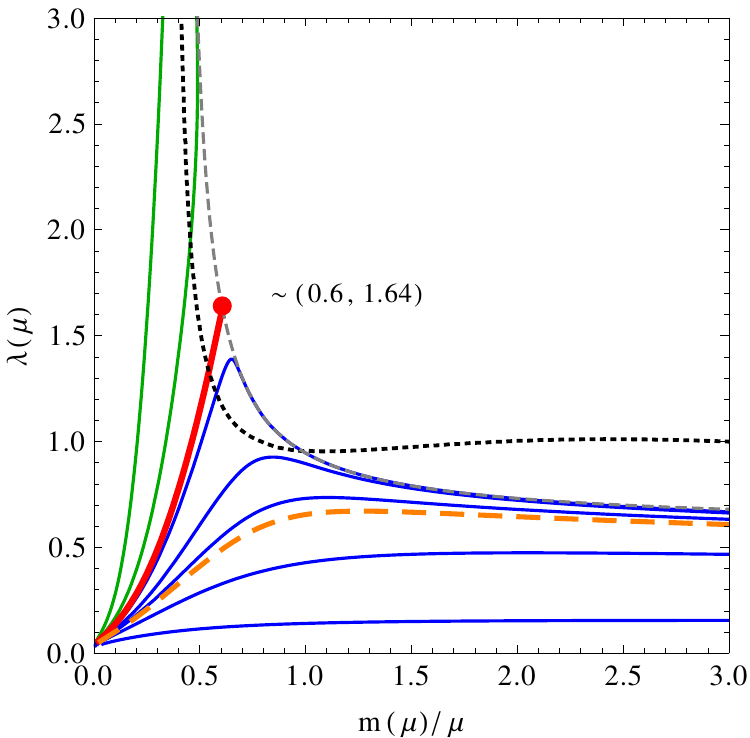}
\caption{Different trajectories of the IS-flow in the plane $(\tilde m\equiv m/\mu,\lambda)$ including two loop corrections.
The infrared fixed point is located at $(\lambda^*,\tilde m^*)\sim (1.64, 0.6)$.
Green trajectories (left side) present a Landau pole, while blue trajectories 
(right side) are infrared safe. The dotted line represent the position for the
extrema of the gluon propagator. The orange (dashed) curve is the one used to 
reproduce lattice data in $SU(3)$.}
\label{Fig:SU3_flujo}
\end{figure}

\begin{figure}
\includegraphics[width=0.45\textwidth]{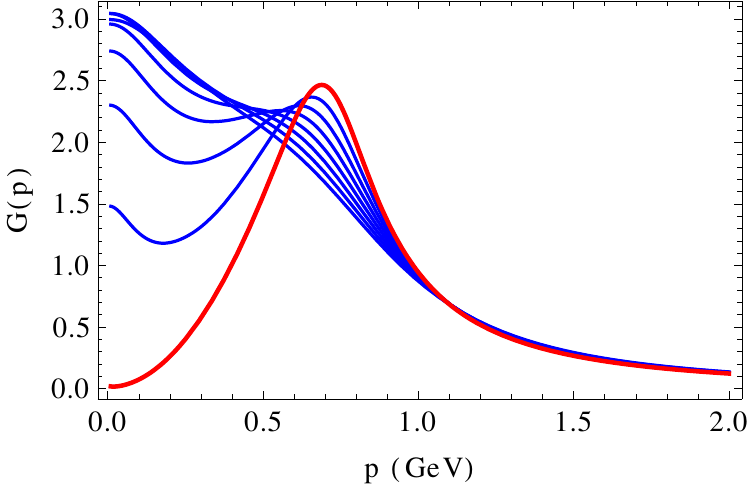}
\caption{Gluon propagator for different initial conditions of the flow.}
\label{Fig:gluonParametrico}
\end{figure}

In Fig.~\ref{Fig:SU3_flujo} we depict the solutions of Eqs. (\ref{Eq:RGflow})
for different initial conditions. The two-loop IR-safe flow presents an 
infrared fixed point at $(\lambda^*,\tilde m^*)\sim (1.64, 0.6)$, where 
$\tilde m=m/\mu$ is the dimensionless mass. These fixed-point values are 
smaller than those found at one loop  $(\lambda^*,\tilde m^*)\sim (16,3.7)$ 
\cite{Reinosa:2017qtf}. The trajectory from the UV fixed point, 
$(\lambda,\tilde m)\sim (0, 0)$, to the infrared fixed point, corresponds to 
the so-called ``scaling solution''. This solution leads to a gluon propagator 
which vanishes at small momentum, as is shown in 
Fig.~\ref{Fig:gluonParametrico}. For initial conditions lying on the left of 
the separatrix, the RG flow terminates at a Landau-pole (green curves), while
the flows initialized on the right of the separatrix (blue trajectories) are 
infrared safe and correspond to a gluon propagator which saturates at a nonzero
value in the infrared, see Fig.~\ref{Fig:gluonParametrico}.

There is a feature which appears at two loops that was not present in our 
previous one loop calculation. We see that, for trajectories close enough to 
the scaling solution, the gluon propagator shows oscillations, see 
Fig.~\ref{Fig:gluonParametrico}. To understand more clearly this phenomenon, we
have drawn in Fig.~\ref{Fig:SU3_flujo} the position of the extrema of the gluon 
propagator (dotted line). This curve corresponds to the values of $\lambda$ and
$m/\mu$ for which the derivative of the gluon propagator with respect to $p^2$ 
is zero, namely:
\[\frac{\beta_\lambda}{\lambda}-\frac{2\mu^2+m^2}{\mu^2+m^2}\frac{\beta_{m^2}}{m^2}+\frac{2\mu^2}{\mu^2+m^2}=0\,.\]

As some of the flow trajectories can intersect this line several times, the 
gluon propagator may present some oscillations, as seen in 
Fig.~\ref{Fig:gluonParametrico}. However, these features concern only
relatively large values of $\lambda$ ($\lambda\gtrsim 1$), a region where the 
perturbative approach is questionable. The oscillations observed in the gluon 
propagator are probably to be attributed to the use of the perturbative 
approach beyond its range of validity. We note that such oscillations are not 
observed in nonperturbative approaches, see eg ref. \cite{Cyrol:2016tym}. Finally, although not shown in Fig.~\ref{Fig:SU3_flujo}, we note that the line of extrema (dotted line) always crosses the IR safe trajectories deep in the infrared.

\subsection{Fixing the parameters}

Our calculation of the gluon and ghost propagator involves four parameters: the
initial conditions of the RG flow (at $\mu_0=1$ GeV) for the running gluon mass
and the running coupling as well as a global normalization of the propagators.

\begin{figure}
\includegraphics[width=0.22\textwidth]{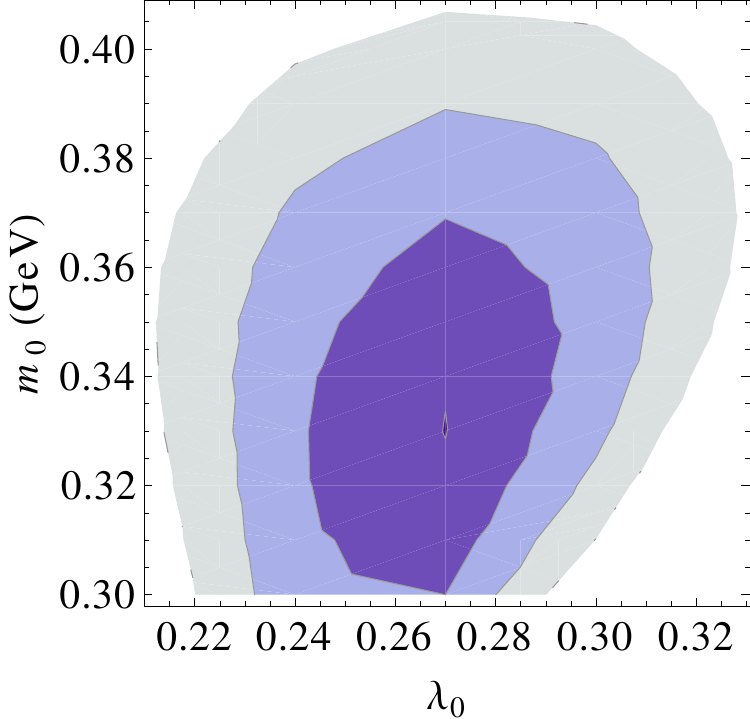}
\includegraphics[width=0.22\textwidth]{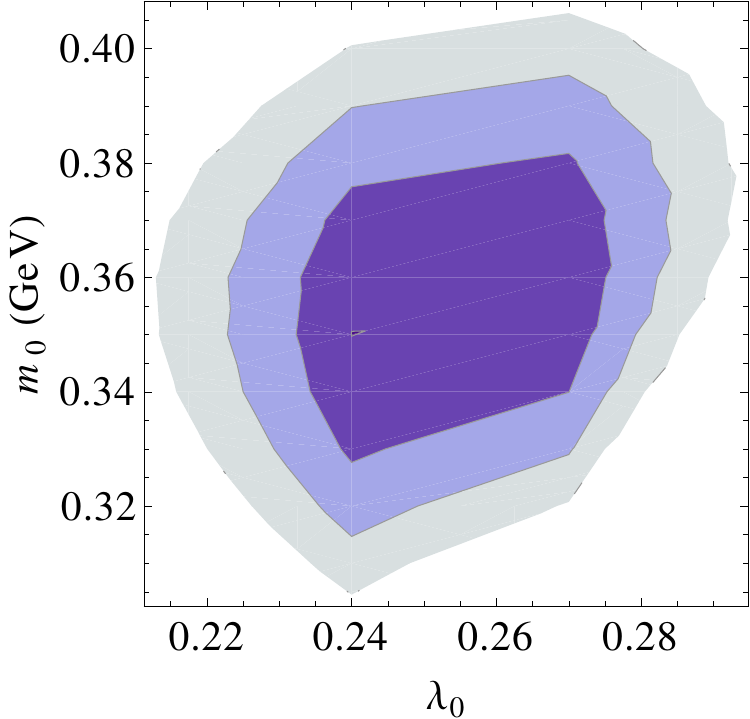}\\
 \includegraphics[width=0.22\textwidth]{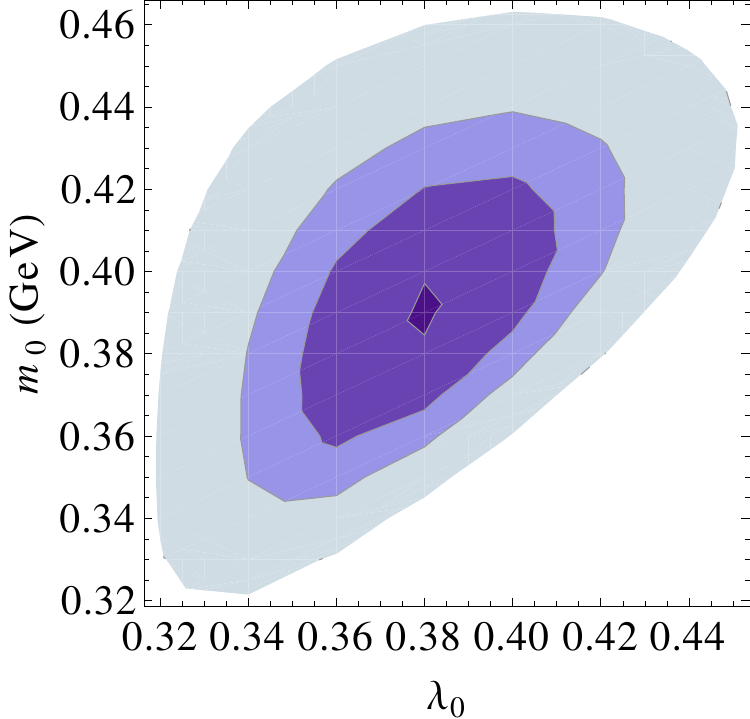}
\includegraphics[width=0.22\textwidth]{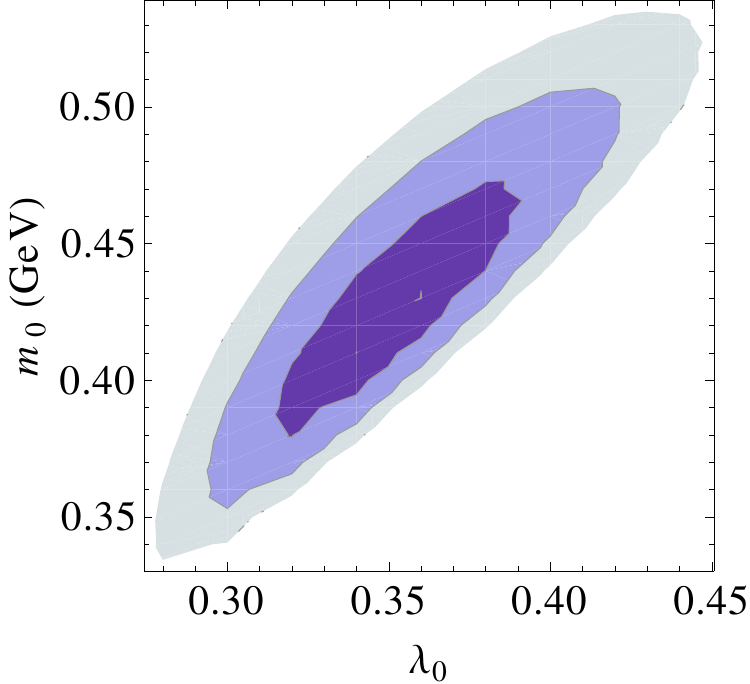}
\caption{Level curves for the error $\chi$ for the two loop correction (left) 
and one loop correction (right) in $SU(3)$ (top) and $SU(2)$ (bottom) all in 
$d=4$ and renormalized using the IS scheme. From dark to light, the different 
regions correspond to 4\%, 5\% and 6\% (top left); 7\%, 8\%, 9\% and 10\% 
(top right); 6\%, 7\% and 8\% and 10\% (bottom left); 7\%, 8\%, 9\% and 10\% 
(bottom right)}
\label{Fig:erroresN3}
\end{figure}

We choose those parameters in order to minimize simultaneously the error for 
the ghost dressing function and the gluon propagator. Specifically, we minimize
the average of the error functions, 
$\chi=\sqrt{\frac{\chi^2_{AA}+\chi^2_{c\cb}}{2}}$, where $\chi^2_{AA}$ and $\chi^2_{c\cb}$ are defined as:
\begin{align}
\label{eq_chi}
\chi^2_{AA}&=\sum_i\frac{G_{\rm lt.}^{-2}(\mu_0)+G_{\rm lt.}^{-2}(p_i)}{2N}\left(G_{\rm 
lt.}(p_i)-G_{\rm th.}(p_i)\right)^2\,,\nonumber\\
\chi^2_{c\cb}&=\sum_i\frac{F^{-2}_{\rm lt.}(\mu_0)+ F^{-2}_{\rm lt.}(p_i)}{2N}
\left(F_{\rm lt.}(p_i)-F_{\rm th.}(p_i)\right)^2\,.\nonumber\\
\end{align}
The subscript ${\rm lt.}$ indicates the lattice data while ${\rm th.}$ 
indicates the perturbative results and $N$ represents the number of 
lattice points with momentum less than $4$ GeV (more  ultraviolet data were disregarded). Therefore, the functions $\chi$
correspond to a sort of average between the (normalized) absolute error and the
relative error. As an example, for the determination of the best fitting 
parameters, we show in Fig.~\ref{Fig:erroresN3} the level curves for $\chi$ as a
function of $\lambda_0$ and $m_0$, for one and two loop corrections, using the 
corresponding data for $SU(3)$ \cite{Duarte:2017wte} and $SU(2)$ 
\cite{Cucchieri:2008qm}. We stress that the optimal value for the coupling 
constant $\lambda_0$ is smaller in the $SU(3)$ than in $SU(2)$ case.
In table \ref{Table:parameters} we summarize the values of the parameters 
$\lambda_0=\lambda(\mu_0)$ and $m_0=m(\mu_0)$ used for the comparison with 
lattice data for different gauge groups. The values of the gluon masses are comparable with those found by other methods, see, eg \cite{PhysRevD.97.034010}. A more quantitative comparison would require using the same renormalization scheme.

\begin{table}[htp]
\begin{tabular}{|l|c|c|c|c|c|c|}
 \hline
$SU(N)$ & \multicolumn{3}{| c |}{Two-loops} & \multicolumn{3}{| c |}{One-loop}\\
\hline
 \hline
  & $\lambda_0$ & $m_0$ (GeV) & \hspace{1mm}$ \chi $ \hspace{1mm}& $\lambda_0$ & $m_0$ (GeV) & \hspace{1mm}$ \chi $\hspace{1mm} \\
\hline
\hline
 $SU(3)$ & 0.27 & 0.33 & 4\% & 0.24 & 0.35 & 7\% \\
\hline
 $SU(2)$ & 0.38 & 0.39 & 6\% & 0.36 & 0.43 & 7\% \\
 \hline
\end{tabular}
\caption{\label{Table:parameters}Parameters used in our calculations with the 
IS scheme, which correspond to the minimum of the error $\chi$.}
\end{table}

\subsection{$SU(3)$}

\begin{figure}
 \includegraphics[width=0.45\textwidth]{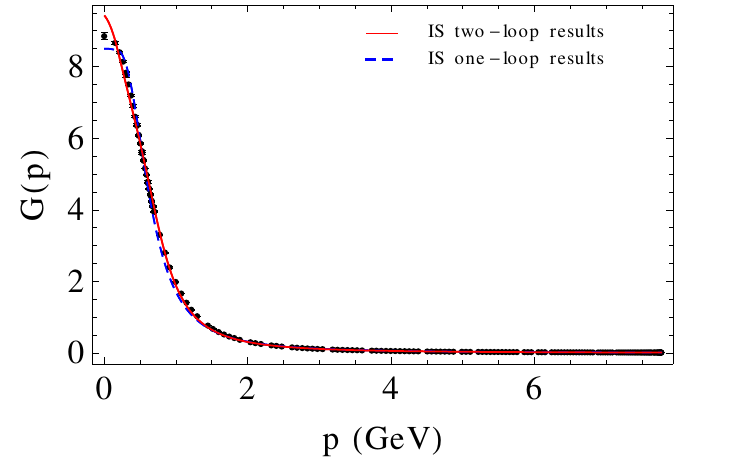}\\
 \includegraphics[width=0.45\textwidth]{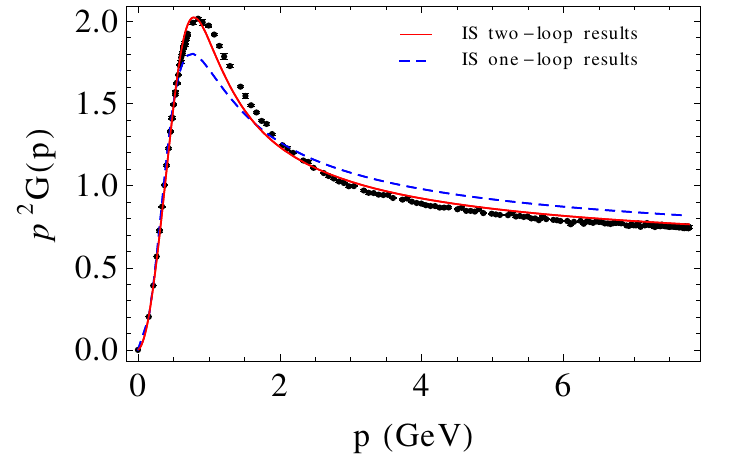}\\
 \includegraphics[width=0.45\textwidth]{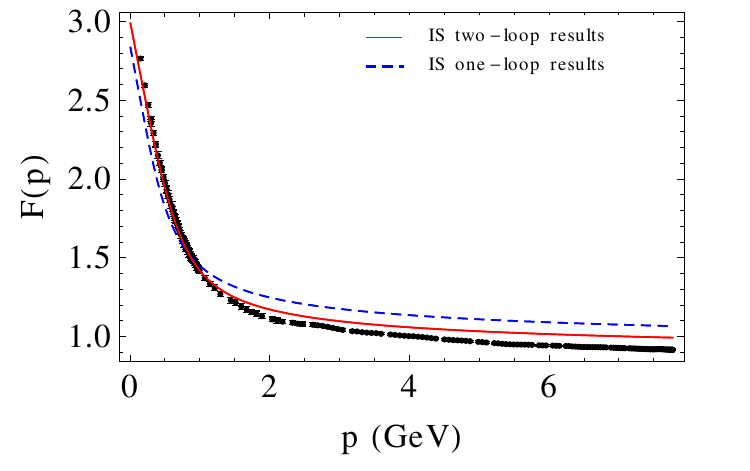}
\caption{Comparison with lattice data from \cite{Dudal:2018cli} for the gluon 
propagator (top), gluon dressing function (middle) and with lattice data from \cite{Duarte:2017wte} for the ghost dressing function 
(bottom) in four dimensions and for the SU(3) gauge group.}
\label{Fig:SU3_IS_2y1L}
\end{figure}

With the set of parameters given in Table~\ref{Table:parameters}, we obtain the
propagators depicted  in Fig.~\ref{Fig:SU3_IS_2y1L}, where we represent the 
two-loop results for the gluon propagator, gluon and ghost dressing functions 
in comparison with lattice data and the one loop calculation. We include the plots for both the gluon propagator and the gluon dressing function since the first one 
yields a better comparison in the deep infrared while the second has better 
resolution for intermediate momenta. 

The agreement between lattice data and perturbation theory is considerably 
improved when the two-loop corrections are added and the error is reduced by a
factor 2. For the two loop results, the error is less than 5\%. At a more 
qualitative level, we observe that the two loop results reproduce the gluon 
dressing function with great accuracy while providing a better fit for the 
ghost propagator.

\begin{figure}
\includegraphics[width=0.48\textwidth]{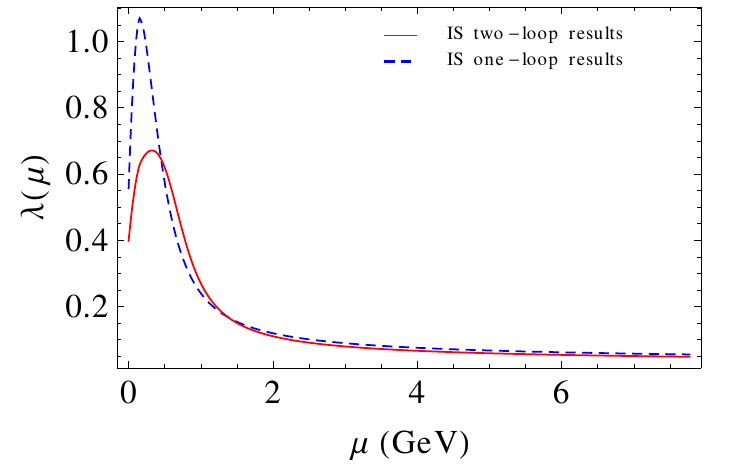}
 \includegraphics[width=0.48\textwidth]{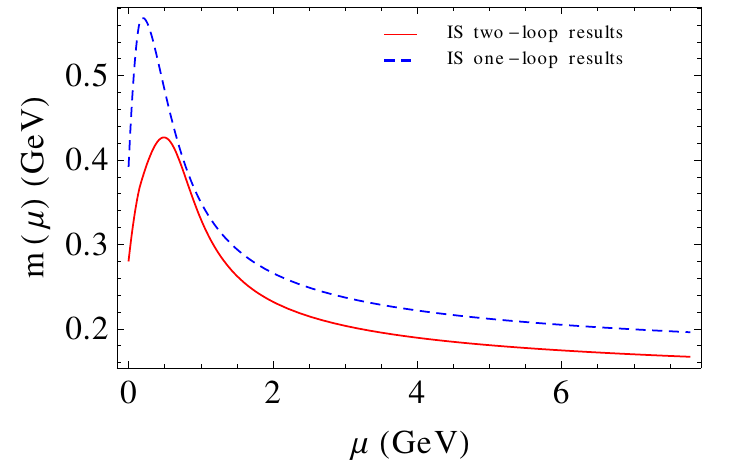}
\caption{Renormalization-group flow of the coupling constant (top) and mass 
(bottom) for the $SU(3)$ gauge group, in the IS scheme.}
\label{Fig:SU3_alphaMasa}
\end{figure}

\begin{figure}
\includegraphics[width=0.45\textwidth]{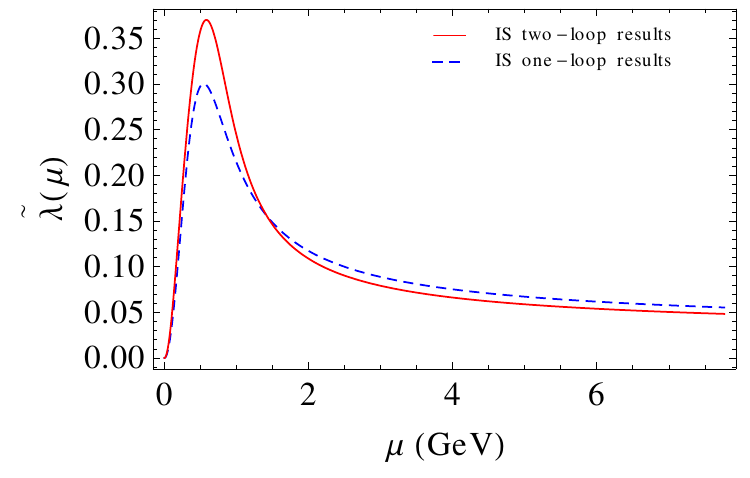}
\caption{Expansion parameter $\tilde\lambda(\mu)$ 
[See Eq.~(\ref{eq_lambdatilde})] as a function of the renormalization-group 
scale for $SU(3)$.}
\label{Fig:SU3_relevant_alpha}
\end{figure}

In Fig.~\ref{Fig:SU3_alphaMasa}, we represent the running gluon mass 
and the squared coupling constant $\lambda$ as a function of the momentum 
scale. 
This figure shows that there is a sizeable difference between the one loop and 
two loop results, but only in a rather small range of momentum. Moreover, as 
discussed in \cite{Reinosa:2017qtf} the relevant expansion parameter within 
this model is not $\lambda$ itself. Indeed  in the deep infrared all the 
interactions are mediated by a gluon propagator, which is massive. A better 
measure of the relative importance of the different terms in perturbation 
theory is 
\begin{equation}
\label{eq_lambdatilde}
    \tilde\lambda(\mu)= \frac{g^2 N_c}{16\pi^2} \frac{\mu^2}{\mu^2+m^2(\mu)}\,.
    \end{equation}
We can see, in Fig.~\ref{Fig:SU3_relevant_alpha}, that the relevant expansion 
parameter does not change much from one loop to two loops and in particular 
it remains less that $0.4$.
It is interesting to observe that the typical error of the 1-loop and $2$-loop 
calculations are not too far from $0.4^2$ and $0.4^3$, which gives a strong 
indication that $\tilde \lambda$ is indeed a good measure of the convergence of
perturbation theory.

\begin{figure}
\includegraphics[width=0.42\textwidth]{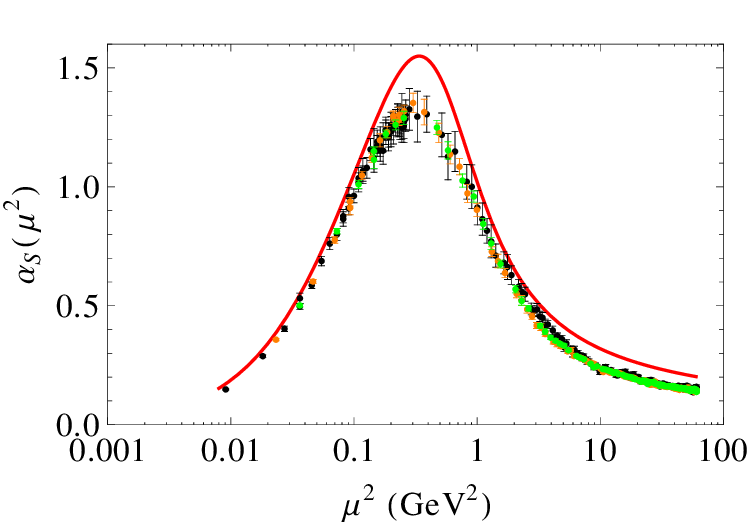}\\
\includegraphics[width=0.42\textwidth]{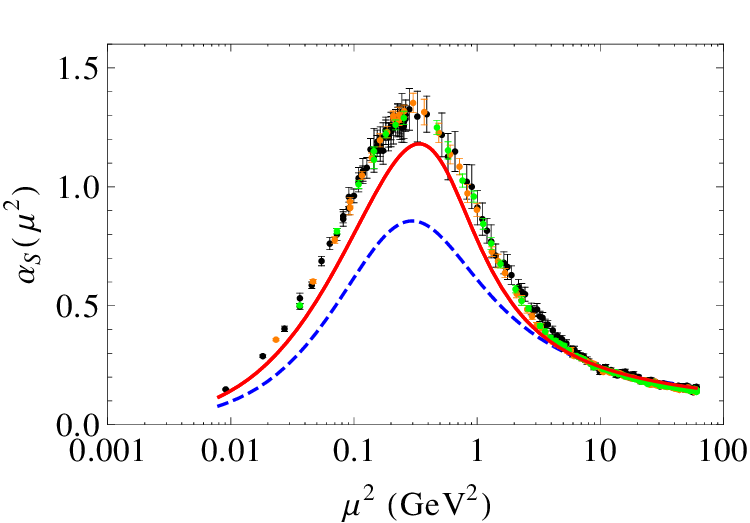}
\caption{The strong coupling constant in the Taylor scheme. The data points are those of \cite{Duarte:2016iko}. The dashed and plain lines correspond to our one-loop and two-loop results respectively. In the lower plot, a correction factor is applied to the value of $\alpha_S$ in the UV, see the main text for more explanations.}
\label{fig:alpha_S}
\end{figure}

We mention finally that, as discussed in Ref.~\cite{Reinosa:2017qtf}, $\tilde\lambda(p)$ corresponds in fact to the Taylor coupling\footnote{One should rewrite $\lambda_0 G(p,\mu_0)$ as $\tilde\lambda_0 (1+\tilde m^2(\mu_0))G(p,\mu_0)$ where $\tilde\lambda_0$ is the Taylor coupling at the initial UV scale $\mu_0$ and $(1+\tilde m^2(\mu_0))G(p,\mu_0)$ is the propagator in the Taylor scheme, equal to $1/\mu_0^2$ when $p=\mu_0$.}
\begin{equation}
\tilde\lambda(p)=\lambda_0\,p^2 G(p,\mu_0)F^2(p,\mu_0)\,,
\end{equation}
which is used in both lattice and continuum studies \cite{Bogolubsky09,Fischer08}. Up to multiplication by a factor of $4\pi/N_c$, it gives the corresponding $\alpha_S(p)$ which we display in Fig.~\ref{fig:alpha_S} at one- and two-loop order in our approach, as compared to the lattice results of \cite{Duarte:2017wte}. It is to be noted that a direct comparison (upper plot) is not really meaningful since the Taylor coupling involves not only the gluon and ghost propagators but also the value of the running coupling at the initial UV scale $\mu_0$. Even though we are able to find good fits for the propagators, there is no reason why the exact value of the coupling in the UV would be reproduced by our fixed loop order calculations. For this reason, we have also considered implementing a correction factor `a' to be applied to $\alpha_S(\mu_0)$ (disregarding possible uncertainties from the lattice). When implementing this recipe, we find the result in the lower plot of Fig.~\ref{fig:alpha_S}, which shows some apparent convergence, with a correction factor `a' changing from the value $0.68$ at one loop to the value $0.76$ at two loops (a correction factor equal to one meaning that no correction factor needs to be applied). We note also the qualitative agreement of our results with the holomorphic reconstruction of $\alpha_S$ put forward in \cite{Ayala:2017tco}.\footnote{We thank Gorazd Cvetic for pointing out this reference to us.}

\subsection{$SU(2)$}
\begin{figure}
\includegraphics[width=0.45\textwidth]{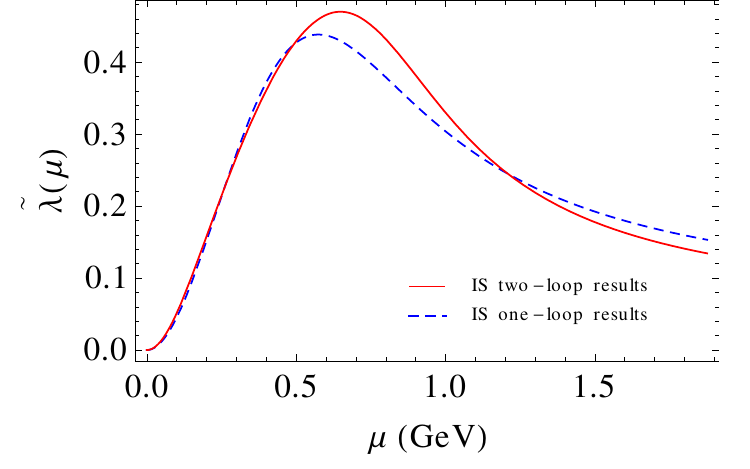}
\caption{Expansion parameter $\tilde\lambda(\mu)$ 
[See Eq.~(\ref{eq_lambdatilde})] as a function of the renormalization-group 
scale for $SU(2)$.}
\label{Fig:SU2_relevant_alpha}
\end{figure}

\begin{figure}
 \includegraphics[width=0.45\textwidth]{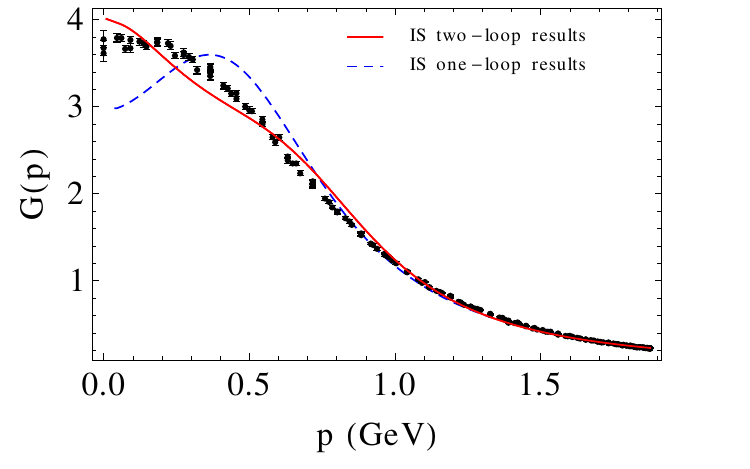}
 \includegraphics[width=0.45\textwidth]{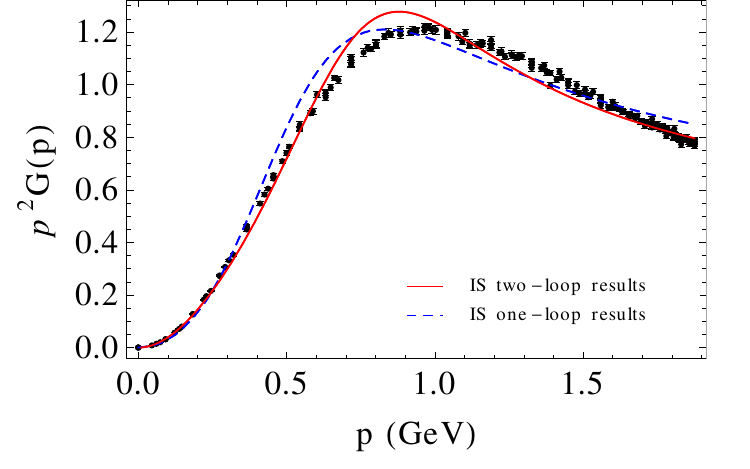}
\includegraphics[width=0.45\textwidth]{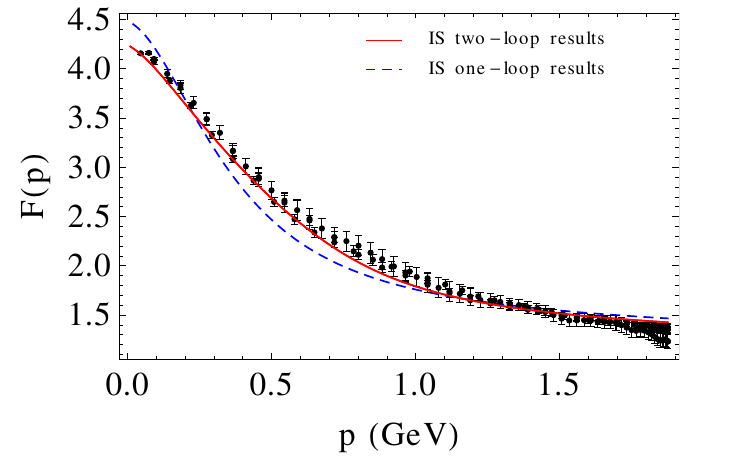}
\caption{Comparison with lattice data from \cite{Cucchieri:2008qm} for the gluon propagator (top), gluon dressing function (middle) and ghost dressing function (bottom)
in four dimension and $SU(2)$.}
\label{Fig:SU2_IS_2y1L}
\end{figure}

We can easily extend our study to the $SU(2)$ case. The one loop case was 
studied in \cite{Tissier:2011ey} and it was observed that the agreement with 
lattice data was less satisfactory than for $SU(3)$. This can be understood 
because the coupling constant seems to be roughly 30\% larger than for $SU(3)$,
as can be seen in Fig.~\ref{Fig:SU2_relevant_alpha}. Using two loops 
corrections, we find parameters that give accurate results for both propagators 
at the same time. Those parameters considerably improve the fitting for gluon 
and ghost dressing functions as is shown in Fig.~\ref{Fig:SU2_IS_2y1L}.

\subsection{Scheme dependence}

The IR safe renormalization-group scheme that we used in this study has several
nice properties but is just one scheme among infinitely many. Of course, in an 
exact calculation, the physical results would not depend on the choice of 
scheme but this property is not maintained when we perform an approximation, 
such as a perturbative expansion.  

In \cite{Tissier:2011ey,Pelaez:2013cpa} we studied the dependence of the 
Yang-Mills propagator with the renormalization scheme. We compared the results 
obtained in the IS scheme with the those obtained in the vanishing-momentum 
scheme (VM), which differs from the IS scheme by changing the condition of 
Eq.~(\ref{IScondition}) by 
\begin{align}
G(p=0)=\frac{1}{m^2}.
\end{align}
This renormalization scheme is not infrared safe so it reaches a Landau pole at
low momentum. As a consequence, we cannot evaluate, as in the IS scheme, 
equation (\ref{eq_int_RG}) at the renormalization scale $\mu=p$. Instead, we 
use $\mu= \sqrt{p^2+\alpha m_0^2}$ where $\alpha$ is a constant. This choice is
sufficient to avoid large logarithms. The conclusion of this study was that the
difference between the results obtained in the IR scheme and the VM scheme were
of the same order as the difference between lattice simulation and the one-loop
calculation. 

In this section, we perform again this comparison with our two-loop results. We
will use the values $\alpha=1$ and $\alpha=2$. The optimal parameters for the 
coupling constant and the mass are presented in Table 
\ref{Table:parametersVM}.
\begin{table}
\begin{tabular}{|l|c|r|}
 \hline
 Two-loops & $\lambda_0$& $m_0$ (GeV)\\
\hline
 VM $\alpha=1$& 0.39 & 0.50\\
 VM $\alpha=2$& 0.36 & 0.50 \\
\hline
  \hline
 One-loop & $\lambda_0$& $m_0$ (GeV)\\
\hline
 VM $\alpha=1$&  0.36 & 0.50\\
 VM $\alpha=2$&  0.42 & 0.50\\
  \hline
\end{tabular}
\caption{\label{Table:parametersVM}Parameters used in our calculations in the 
VM scheme in $SU(3)$, which  minimize the error $\chi$.}
\end{table}

\begin{figure}
\includegraphics[width=0.45\textwidth]{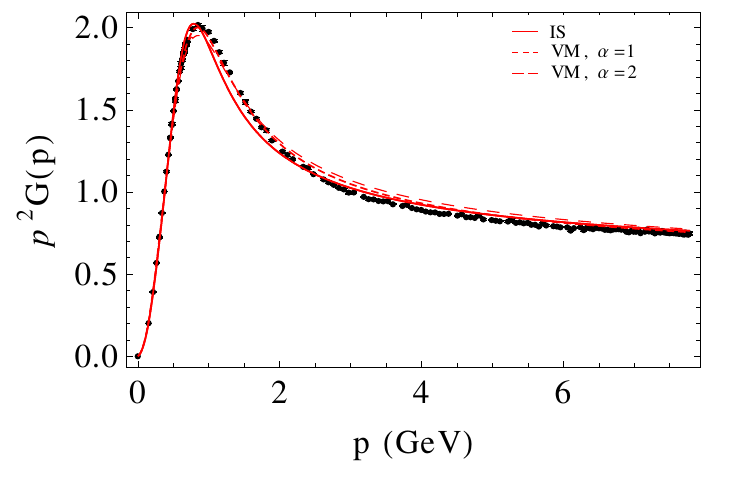}
  \includegraphics[width=0.45\textwidth]{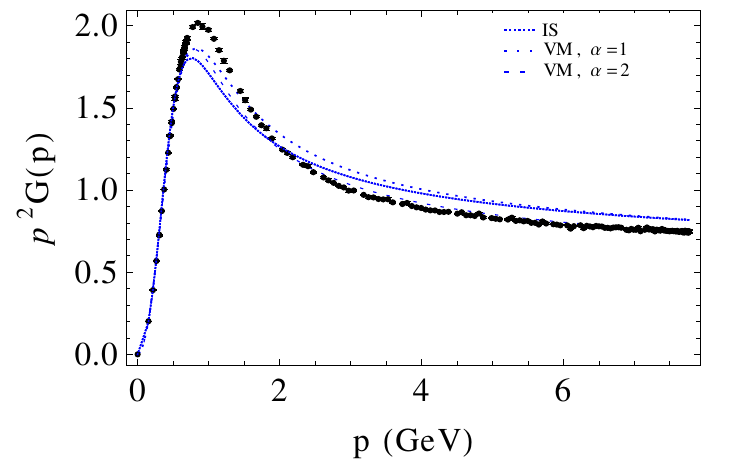}\\
\caption{Comparison with lattice data from \cite{Duarte:2017wte} for the gluon dressing function 
in four dimension and $SU(3)$ using different schemes at two loops (top) and one loop (bottom).}
\label{Fig:SU3_schemes_2L}
\end{figure}
In Fig.~\ref{Fig:SU3_schemes_2L}, we analyze the dependence in the 
renormalization scheme by comparing our results computed in the IS scheme, and 
in the VM scheme for $\alpha=1$ and $\alpha=2$. We can see that the dependence 
on the scheme for one loop results in $SU(3)$ is small. Still, as expected, two
loop results reduce this dependence as it is shown in 
Fig.~\ref{Fig:SU3_schemes_2L}.

In order to measure quantitatively the difference between schemes we compute 
the error with respect to the IS scheme, $\mathcal{H}$, defined as:
\begin{equation}
\mathcal{H}(\alpha)=\sqrt{\frac{1}{N}\sum_i\frac{\left(\eta_{\rm VM(\alpha)}(p_i)-\eta_{\rm IS}(p_i)\right)^2}{\eta^2_{\rm IS}(p_i)}}\nonumber\\
\end{equation}
where $\eta$ represents the gluon or ghost dressing functions and $N$ the 
number of lattice points. In Table \ref{Table:schemesN3}, we summarize the 
values of $\mathcal{H}$ for gluon and ghost dressing functions using one loop 
or two loops results.

\begin{table}
\begin{tabular}{|l|c|c|c|r|}
\hline
& \multicolumn{2}{| c |}{Gluon dressing} & \multicolumn{2}{| c |}{Ghost dressing}\\
\hline
   & One-loop & Two-loops & One-loop & Two-loops\\
  \hline
  VM $\alpha=1$ & 0.03 & 0.02 & 0.06 & 0.05\\
  \hline
  VM $\alpha=2$ & 0.06 & 0.03 & 0.08 & 0.03\\
\hline
\end{tabular}
\caption{\label{Table:schemesN3}Estimate of the scheme dependence $\mathcal{H}$ between the IS scheme and  different VM schemes in $SU(3)$ for the gluon and ghost dressing functions.}
\end{table}

\section{Conclusions}
\label{sect_conclusion}
In this article, we have computed the propagators of the gluons and ghosts at 
two-loop order in the Curci-Ferrari model, in the quenched approximation. These were
compared with the available lattice simulations, both for $SU(2)$ and $SU(3)$ 
gauge groups. The gluon mass is seen as a phenomenological parameter, which is 
fitted to obtain the best agreement with the lattice data. The two loop 
calculations significantly improve the fits for the $SU(3)$ group. With a 
unique set of fitting parameters, we obtain a maximal error of a few percent on 
the gluon and ghost propagators. In the $SU(2)$ case, we also find an 
improvement of the precision, but which is less significant. This can be traced
back to the fact that the interaction is bigger in $SU(2)$ than in $SU(3)$. 

This study gives strong indications that the Curci-Ferrari model is indeed a good 
phenomenological model for describing the correlation functions of QCD in the 
quenched approximation. We stress that it is a nontrivial result that two-loop 
results reproduce better lattice simulations than the corresponding ones at one-loop order. Indeed, it could 
happen that, adding more and more loops, we obtain correlation functions that 
converge to results very different from the lattice results. This possibility 
seems to be excluded by our analysis. Moreover, it justifies {\it a posteriori}
that a good estimate of the infrared contributions of higher loops is given by 
the square of the coupling constant, divided by the typical momentum squared 
[see Eq.~(\ref{eq_lambdatilde})] (up to multiplicative factors). This is an 
important effect, which ensures the convergence of perturbation theory in the 
deep infrared regime.

Following the same procedure, a calculation of the quark propagator could be 
performed. The situation is more complex in this case because two masses are present, those of the quark and the gluon, leading to an increase in the number of master integrals appearing in the expressions. This 
calculation is interesting because the renormalization factor of the quarks 
receives no correction at 1-loop in the Landau gauge. In this situation, we 
expect the 2-loop contribution to have a significant influence on the results. 
For higher point vertices, the calculations become significantly more complex.
In particular, while the Laporta algorithm will decompose Feynman diagrams to 
master integrals, full analytic expressions for massive two loop $n$-point 
masters are not known. This could be circumvented by considering $n$-point 
vertices with $n-2$ vanishing external momenta.

Alternatively, one could compute power corrections, such as $\frac{m^2}{p^2}$,
to the full two loop vertex functions with non-nullified external legs in order
to compare with lattice results in intermediate momentum ranges. This would
provide an interim study of when mass effects become apparent ahead of when the
fully two loop technology becomes available.

\begin{acknowledgments}
We would like to thank Orlando Oliveira for kindly accepting to share data on the gluon and ghost correlators as well as on the running coupling. J. A. G. gratefully acknowledges CNRS for a Visiting Fellowship and the hospitality of LPTMC, Sorbonne University, Paris, where part of the work was carried out as well as the support of the German Research Foundation (DFG) through a Mercator Fellowship. M. P. and U. R. acknowledge support from ANII-FCE (``Agencia Nacional de Investigación e Inovación - Fondo Clemente Estable''), Grant No. 126412. M.P. and M. T. acknowledge support from ECOS program, Grant No. U17E01. Figures 1 and 2 were drawn with the use of {\sc Axodraw}, \cite{jag11}.
\end{acknowledgments}

\end{document}